\documentclass[man]{article}
\usepackage{amsmath, amsthm, amsfonts}
\usepackage{apacite}
\usepackage{graphicx}
\usepackage{epsfig}
\usepackage{pstricks}
\usepackage{pst-text}
\usepackage{pst-node}
\usepackage{subfigure}

\newtheorem{prop}{Proposition}
\newtheorem*{prop1}{Proposition 1}
\newtheorem{assumption}{Assumption}

\begin{document}

\title{Minimal Assumption Derivation of a weak Clauser-Horne Inequality}
\author{Samuel Portmann\thanks{Institute of Theoretical Physics/History and
    Philosophy of Science, Sidlerstrasse~5, University of
    Bern, CH-3012 Bern, Switzerland.}, Adrian W\"uthrich\thanks{History and
    Philosophy of Science, Exact Sciences, Sidlerstrasse~5, University of
    Bern, CH-3012 Bern, Switzerland.}}

\date{\today}
\maketitle
\begin{abstract}
According to Bell's theorem a large class of hidden-variable models
obeying Bell's notion of local causality conflict with the predictions of quantum mechanics. Recently, a Bell-type
theorem has been proven using a weaker notion of local causality, yet assuming the existence of
perfectly correlated event types. Here we present a
similar Bell-type theorem without this latter assumption. The derived inequality differs from the Clauser-Horne inequality by some
small correction terms, which render it less
constraining.
\vspace{0.5cm}
\end{abstract}
\tableofcontents

\section{Introduction}
In this article we continue the work of
\citeA{grasshoff05}\footnote{See also \citeA{wuethrich04}.} and prove
a Bell-type theorem from a still weaker set of assumptions. In
contrast to \citeA{grasshoff05}, the weakening is reflected in the derived inequality: We get the Clauser-Horne inequality with
small correction terms rendering our inequality less
constraining.\footnote{A very similar result for the special
  case of two-valued common causes was derived independently by \citeA{hofer06}.}

\begin{figure}
\begin{center} 
\epsfig{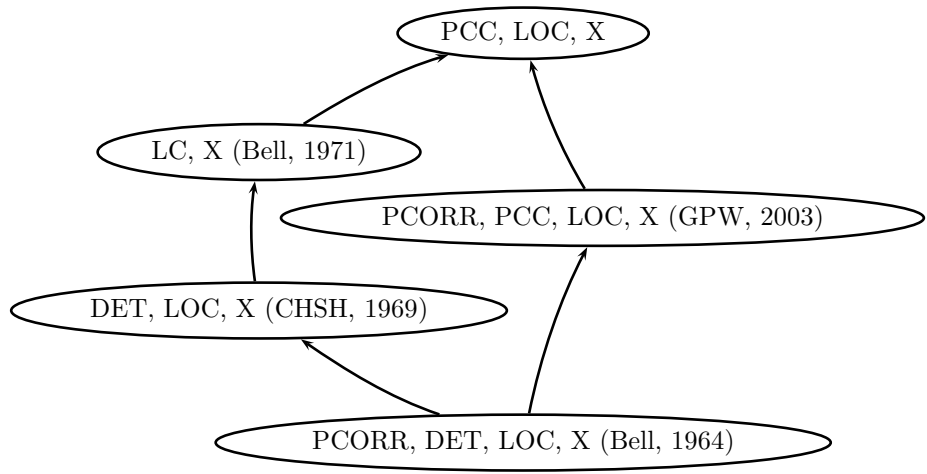}
\caption{\label{fig1} Comparison of the logical strengths of the
  different sets of assumptions. Each node stands for a set of
  assumptions from which a Bell-type inequality was derived. If two nodes are
  connected by an arrow, the set further upwards is a logical implication of
  the set further downwards but not vice versa. Caption: DET=Determinism,
  LOC=Locality, PCORR=$\exists$ perfectly correlated event types, LC=local causality, PCC=Reichenbach's principle of common cause, X=further assumptions, shared by all derivations.}
\end{center}
\end{figure}

There are many different Bell-type theorems with different aims, the
weakening of the assumptions being one among many
objectives.\footnote{E.g.~GHZ-type theorem's (\citeA{greenberger89}) foremost achievement is simplicity.}
In order to set the theoretical stage, we would like to recall some works\footnote{For more
  detailed reviews see e.g.~\citeA{shimony05} and \citeA{clauser78}.}
aiming to minimalize the strength of the assumptions and set
them in context to our own work (see figure \ref{fig1}).

The experimental context of all these derivations is the EPR-Bohm
experiment (see section \ref{eprb}). Furthermore, they all assume a
locality and a
causality condition\footnote{By
  choosing this terminology, we do however not intend to exclude the
  possibility that there can still be non-local causality, even if
  only the causality condition is violated (see e.g.~\citeA{butterfield92},
  \citeA{butterfield922}, \citeA{jones93}, and \citeA{maudlin94}).} for the observable events in terms of ``hidden'' variables. In its
canonical interpretation, quantum mechanics (QM) violates the locality but not the
causality condition.
In his seminal derivation, \citeA{bell64} assumed local determinism
(LOC and DET) and, additionally, the existence of perfectly correlated event types
(PCORR). Then, \citeA{clauser69} derived the CHSH-inequality---again
with LOC and DET, but without PCORR. Moreover, \citeA{bell71} showed two years
later, that the same inequality can even be derived if one replaces the
assumption of local determinism
 with a weaker probabilistic notion, which he dubbed ``local
causality'' (LC) \cite{bell76}, and which was later analyzed by \citeA{suppes76}, \citeA{vanfraassen82} and
\citeA{jarrett84} as a conjunction of a
locality and a causality condition. As the long philosophical discussion
demonstrates, it is already difficult to find an only necessary condition
for probabilistic causation. And already \citeA{bell76} stressed that other
definitions of LC are conceivable.  \citeA{belnap96} and \citeA{redei99}
showed that Reichenbach's Principle of Common Cause does indeed suggest
another form of causality (PCC), which, together with LOC, includes Bell's notion only as a special
case. They also pointed out, that the existing proofs all assume
the stronger notion and that it is thus not clear
whether a Bell-type theorem can still be proven with PCC. We used PCC
as our causality condition
in \citeA{grasshoff05} (GPW) for a proof of a Bell-type theorem, but
the minimality of the logical strength of the assumptions was only
relative (see figure \ref{fig1}), because we also assumed PCORR. \emph{Given}
this assumption, our set of assumptions was minimal. However, there are
reasons to think that PCORR is false
(see section \ref{maxcorr}), which limits the
significance of our result. In this article, we derive a
Bell-type inequality without assuming PCORR. Our approach is similarly ``straightforward'' as the one of
\citeA{ryff97}. His intuition ``that if a theorem 
is valid whenever we have perfect correlations, it cannot be totally wrong in
the case of almost perfect correlations'' can be formulated precisely
and proven to be correct in our case. 

This article is structured as follows. We
describe the EPRB experiment and introduce our
notation in section \ref{eprb}. In the main part, section \ref{proof},
we derive a weak Clauser-Horne inequality. In section
\ref{discussion}, 
we discuss our result and compare it to related
work. Specifically, we discuss the significance of the small correction
terms in our inequality.

\section{The EPRB experiment}\label{eprb}
\begin{figure}
   \includegraphics[width=\linewidth]{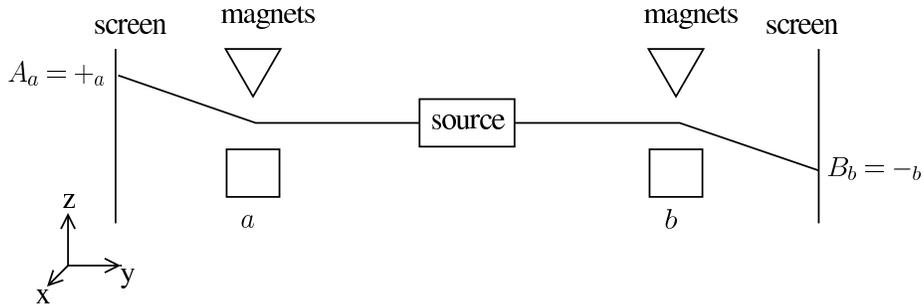}
\caption{\label{fig:eprb}Setup of the EPR-Bohm experiment.}
\end{figure}
Consider the so-called \emph{EPR-Bohm (EPRB) experiment} 
\cite{epr,bohm51}. Two spin-$\frac{1}{2}$ particles in the \emph{singlet state}

\begin{align}
   |\Psi\rangle =\frac{1}{\sqrt{2}}\left( |\!\uparrow\downarrow\rangle
     -|\!\downarrow\uparrow\rangle\right)
\end{align}
are separated such that one particle moves to the measurement
apparatus of Alice on the left and the other particle to the
measurement apparatus of Bob on the right (see figure \ref{fig:eprb}). The
experimenter can arbitrarily choose the direction in which the 
spin is
measured with a Stern-Gerlach magnet.

The event\footnote{If not otherwise stated we refer with ``events'' always
to event \emph{types}, in contrast to event tokens which instantiate types.} that Alice's (Bob's) measurement apparatus is set to measure the spin in
direction~$a$ ($b$) is symbolized by $a$ ($b$).  $A_a$ ($B_b$)
symbolizes the measurement outcome of Alice (Bob) for a measurement 
in direction $a$ ($b$). For each
direction, there are two possible
measurement outcomes: \emph{spin up} ($A_a=+_a$, $B_b\!=\!+_b$) and \emph{spin down} 
($A_a=-_a$, $B_b\!=\!-_b$).\footnote{\label{fn:1}It is sometimes 
 said that the
  assumptions of Bell-type theorems are
  completely independent of QM. However, it seems to us that at least
  the event structure with $A_a$, $B_b$, $a$ and $b$ is adopted
  from QM. Alternatively, one could try to start from scratch
  by invoking an independent criterion for event
  identity. Some steps in this direction have been done by
  \citeA{butterfield922}
 using David Lewis's account of events.} We will always interpret these events also as elements
of a Boolean algebra $\Omega$ with a classical probability measure
$p$, constituting a classical probability space $(\Omega, p)$. E.g.
\begin{align}\label{2a}
   p(A_aB_b|ab)
\end{align}
denotes the probability that Alice' measurement outcome is $A_a$ and
Bob's $B_b$, when measuring in the directions $a$ (Alice) and $b$ (Bob).
We will often use the notation\footnote{Note that, if
 $p(\dots)$ is a probability measure and $p(ab)\neq 0$, then $p_{a,b}(\dots)$
is also a probability measure.}
\begin{align}
p_{a,b}(\dots):=p(\dots|ab),
\end{align}
with which we can write (\ref{2a}) as
\begin{align}
p_{a,b}(A_aB_b).
\end{align}
These probabilities are predicted by quantum mechanics as 
\begin{align}
   p_{a,b}(+_a+_b)&=\frac{1}{2}\sin^{2}\frac{\varphi_{a,b}}{2}, \label{cor1}\\
   p_{a,b}(-_a-_b)&=\frac{1}{2}\sin^{2}\frac{\varphi_{a,b}}{2}, \label{cor2}\\
   p_{a,b}(+_a-_b)&=\frac{1}{2}\cos^{2}\frac{\varphi_{a,b}}{2}, \label{cor3}\\
   p_{a,b}(-_a+_b)&=\frac{1}{2}\cos^{2}\frac{\varphi_{a,b}}{2}, \label{cor4}
\end{align}
where $\varphi_{a,b}$ denotes the angle between the two measurement directions
$a$ and $b$. Also, the outcomes on each side are predicted separately to be 
completely random:
\begin{align}
   p_{a,b}(A_a)=\frac{1}{2}, \label{rand1}\\
   p_{a,b}(B_b)=\frac{1}{2}. \label{rand2}
\end{align}

\section{Proof of a Bell-type theorem}\label{proof}
We will first introduce our assumptions (sections~\ref{assumptions} and \ref{maxcorr}). In the literature
on Bell-type theorems, there is a huge amount of work devoted to the
discussion of the assumptions. We do not intend to contribute to this
discussion here. Rather, we will discuss the new
elements in our set of assumptions. First, we will see that the first
three assumptions imply the existence of a common cause which screens
off the correlations in question. As in
\citeA{grasshoff05}, the new and crucial
thing here is that, in contrast to other derivations, we do not demand a single common cause for all the
different correlations. Second,
since we now have several different common causes, we need to adjust
some assumptions (\ref{loc2} and \ref{no-cons}) to this new situation. The comments on the remaining
assumptions are there for the sake of clarity and not intended to contribute to the ongoing debate.

\subsection{Locality and Causality}\label{assumptions}

The correlation between `heads up' and `tails down' when tossing a coin is
explained by the identity of the instances of the respective events: Every
instance of `heads up' is also an instance of `tails down', and vice versa.
Large spatial separation of coinciding instances of $A_a$ and $B_b$ suggests
that such is not the case in the EPRB setup:\footnote{See also
  footnote~\ref{fn:1}.}

\begin{assumption}\label{sep}
   The coinciding instances of the events $A_a$ and $B_b$ are distinct.
\end{assumption}

Given this assumption, we can express
\begin{assumption}\label{loc1}
   No $A_a$ or $B_b$ is causally relevant for the other. 
\end{assumption}
This assumption is supported by the fact that the measurements can be
made such that in each run of the experiment the instance of $A_a$ is space-like separated
from the instance of $B_b$. If it were
violated and if a cause temporally precedes its effects, the
direction of causation would depend on the chosen inertial
frame.\footnote{Note however that this is \emph{per se} not a
  violation of Lorentz invariance and that whether or not this stands
  in contradiction to the special theory of relativity is an intricate
  matter. For a discussion, see for example \citeA{maudlin94} and \citeA{weinstein06}.}

\begin{assumption}[PCC]\label{pcc}
   If two events $A$ and $B$ with distinct coinciding instances are correlated and neither $A$
   is causally relevant for $B$ nor vice versa, then there exists a
 partition $C=\{C_i\}_{i\in I}$ of $\Omega$, a \emph{common cause}, such 
 that
   \[
   p(A B|C_i)=p(A|C_i)p(B|C_i),\qquad \forall i\in I.
   \]
\end{assumption}

We will assume the cardinality of $I$ to be countable.\footnote{We choose this
  constraint only for simplicity. The derivation can easily be amended also
  for $I$ being uncountable.} The common cause can alternatively be
thought of as a variable (the ``hidden'' variable) taking on the
elements of the partition as values. Thus, when we say ``the value of the common cause'',
we refer to an element of the partition.  In the original formulation,
Reichenbach used a partition with two elements, which is here generalized to a
partition with countably many elements.\footnote{\label{fn:2}\citeA{reichenbach56} and
  \citeA{hofer04} stipulate further conditions, for the two-valued and the
  general case respectively. For our derivation, we do not need these
  assumptions, though.}
\\
Now, as can be seen from equations~(\ref{cor1})-(\ref{rand2}), in general, the event $A_a$
is correlated with event $B_b$:
\begin{align}
   p_{a,b}(A_aB_b)\neq p_{a,b}(A_a) p_{a,b}(B_b),
   \mbox{ except for } \varphi_{a,b}=\frac{\pi}{2} \mod \pi .
\end{align}
With assumptions \ref{sep} and \ref{loc1}, PCC demands the existence
of a common cause $C^{abAB}=\{ C_i^{abAB}\}_{i\in I^{abAB}} $
which screens off the correlation: 
\begin{align}\label{11}
 p_{a,b}(A_aB_b|C_i^{abAB})= p_{a,b}(A_a|C_i^{abAB})
 p_{a,b}(B_b|C_i^{abAB}), \quad \forall i \in I^{abAB}.
\end{align}
As in \citeA{grasshoff05}~there is a common cause $
C^{abAB}$ for
\textit{each} quadruple of measurement directions and outcomes
$(a,b,A_a,B_b)$. That is different from other derivations, where a single
common cause $\{C_i\}_{i\in I}$ is stipulated for all correlated events:
\begin{align}\label{ccc}
p_{a,b}(A_a B_b|C_i)= p_{a,b}(A_a|C_i) p_{a,b}(B_b|C_i), \quad \forall
i\in I.
\end{align}
That such a \textit{common} common cause (obeying (\ref{ccc})) was assumed in 
Bell-type theorems was pointed out and criticized by \citeA{belnap96}
and \citeA{redei99}.

Mathematically, (\ref{ccc}) is stronger
than (\ref{11}). Indeed, as already pointed out by
\citeA{butterfield89} (p.~123), to get from (\ref{11}) to (\ref{ccc}) one
needs at least one further assumption (see also \citeA{grasshoff05},
p.~15 et seq., and \citeA{henson05},
p.~532 et seq.). The additional assumption states that the statistical
independence in (\ref{11}) does not get disrupted if one
(additionally to $C_i^{abAB}$) conditionalizes on other events
which are not causally relevant for $A_a$ and $B_b$. We do not
have any strong argument for or against this assumption. To
be sure, there are arguments which are brought forward in the
literature in favour of it (see e.g.~\citeA{skyrms80},
\citeA{eells83}, and \citeA{uffink99}).\footnote{If one aims at a
  characterization of causality in purely statistical terms, the
  possibility of disrupting causal independencies by causally
  irrelevant factors would be a major obstacle (we thank Michael
  Baumgartner for pointing this out to us). Indeed, one prominent
  approach in this direction, which is based on Bayesian networks
  (\citeA{spirtes93}, \citeA{pearl00}) denies this possibility. But of
  course, its making the sought-after inference from statistics to
causality functional is not an argument for its truth.} But
we think it is fair to say that the issue is contentious (see e.g.~\citeA{cartwright79}). And, since we only need (\ref{11}) for our derivation,
we do not need to take sides on this question.\footnote{Interestingly,
there is a strong argument against this assumption if one embraces the other Reichenbachian conditions (see footnote~\ref{fn:2}) as well. For a given common cause \emph{system} that obeys
the additional assumptions as well, \citeA{hofer04} show that there is no finer partitioning possible
without disrupting the statistical independence in this finer partition.}

\begin{assumption}[LOC]\label{loc2}
\begin{align}
p(A_a| ab C_i^{abAB})&=p(A_a | a C_i^{abAB}), \\
p(B_b | ab C_i^{abAB})&=p(B_b | b C_i^{abAB}).
\end{align}
\end{assumption}
This assumptions is meant to prevent the possibility of superluminal
causation. Rather than to justify LOC, we will recall the
justification of the analogue of it in the traditional derivations
and show that the same justification works in our case as well. In this
way, we back up our minimality claim. 

In the
traditional 
derivations, there is just one common cause. The condition then reads
\begin{align}\label{loc22}
p(A_a | ab C_i)&=p(A_a | a C_i), \quad
\mbox{and} \nonumber \\
p(B_b | ab C_i)&=p(B_b | b C_i) \quad \mbox{(for all
  values of $a$, $b$, $A_a$ and $B_b$)}.
\end{align}
The canonical justification of (\ref{loc22}) runs along the following
lines. One first notes that the EPRB experiment can be set up such
that the measurement outcome $A_a$ ($B_b$) and the choice of the
measurement setting $b$ ($a$) are space-like separated. \emph{If} Alice knew the
value of the common cause (which, say, is part of her past light-cone) and
($\ref{loc22}$) did \emph{not} hold, Bob could send Alice a signal
superluminally by setting up a measurement direction, since this would
alter the corresponding probability of Alice' measurement
outcome. Now we do no longer have just one single common cause for all
correlations but one for each. 
Nonetheless, the justification given above works all the same. In the sentence 
\begin{quotation}
``If Alice knew \emph{the
value of the common cause} (which, say, is part of her past light-cone) and
($\ref{loc22}$) did not hold, Bob could send Alice a signal
superluminally by setting up a measurement direction, since this would
alter the corresponding probability of Alice' measurement
outcome.'',
\end{quotation}
just replace the italics with ``the value of \emph{a} common cause''.\footnote{The
  justification also works if, instead of the value of a single
  common cause, one takes conjunctions or disjunctions of the
  values of common causes, or any element of the subalgebra
  generated by them.} 

\subsection{Common causes for the maximal correlations}\label{maxcorr}

In their derivation \citeA{grasshoff05} exploit that the screening-off
condition entails that common causes of perfect
correlations determine the effects. The slightest
deviation from
\begin{align}\label{perfect}
p_{a=b}(+_a|-_b)=p_{a=b}(+_b|-_a)=1.
\end{align}
leads to a breakdown of that type of derivation. Of course,
equation~(\ref{perfect}) is true according to QM and any apparent violation in
actual experiments may be attributed to
experimental shortcomings, for instance that, in practice, the measurement devices are never set up perfectly parallel. 

Nevertheless, we would like to do without
this assumption. Our motivation for this is twofold.
First, there are theoretical grounds on which to expect a violation of the
quantum mechanical prediction of perfect correlations. Theoretical work
in the different approaches to quantum gravity suggests that tiny violations of Lorentz
group invariance are to be expected.\footnote{See
  e.g.~\citeA{mattingly05} for references.} Seen as an implication of rotation
invariance, (\ref{perfect}) would not be warranted any more. The second motivation has to do with the prominent claim
that Bell-type theorems rule out the existence of empirically adequate
local hidden-variable models on empirical grounds alone. However, if
besides the assumptions that define the model as a local
hidden-variable model, the only constraint were empirical adequacy,
PCORR should not be assumed, because small violations of it are
consistent with empirical data.\footnote{In the context of the
  Kochen-Specker theorem, a similar loophole was exploited to construct a
  non-contextual empirical adequate model by \citeA{clifton00}.}

These considerations motivate a weakening of (\ref{perfect}) such that
we just take the maximal correlations available, without assuming that
they are perfect.
We do
this as follows. For each pair of measurement directions $(a,b)$, we
parametrize the conditional probabilities $p_{a,b}(+_a|-_b)$ and
$p_{a,b}(+_b|-_a)$ as
\begin{align}
p_{a,b}(+_a|-_b)&=1-\epsilon_{a,b}, \nonumber \\
p_{a,b}(+_b|-_a)&=1-\epsilon_{b,a}, \mbox{ with
}\epsilon_{a,b}, \epsilon_{b,a} \in [0,1].
\end{align}
We will call the set of all measurement directions of Alice (of Bob) $D_A$ ($D_B$). For each measurement direction $a\in D_A$
($b\in D_B$), we pick out the measurement direction $\dot{a}\in D_B$ ($\dot{b}\in
D_A$) for which $p_{a,b}(+_a|-_b)$ ($p_{a,b}(+_b|-_a)$) takes on
its maximal value, or, equivalently, $\epsilon_{a,b}$
($\epsilon_{b,a}$) takes on its minimal value.\footnote{If the number
of measurement directions (i.e.~the cardinality of $D_A$ and $D_B$) is
not finite, it is possible, that there is no such minimal value but
only an infimum. The proof can be amended also for this case, but we
will refrain from doing this here.} If the same minimal value
is taken on for more than one direction, we make an arbitrary
choice. We denote this minimal value with $\epsilon_a$
($\epsilon_b$):\footnote{Note that we do not assume that the minimal
value is taken on for parallel measurement directions.}
\begin{align}
\epsilon_a:&=\min_b \{\epsilon_{a,b} \}, \nonumber \\
\epsilon_b:&=\min_a \{\epsilon_{b,a} \}.
\end{align}
Thus, we have 
\begin{align}
p_{a,\dot{a}}(+_a|-_a)&=1-\epsilon_a, \nonumber \\
p_{\dot{b},b}(+_b|-_b)&=1-\epsilon_b.
\end{align}

Because of assumptions \ref{sep} to \ref{pcc}, we have (in the notation of formula (\ref{11})) a common cause
$\{C_i^{a\dot{a}+-}\}_{i\in I^{a\dot{a}+-}}$
($\{C_i^{\dot{b}b-+}\}_{i\in I^{\dot{b}b-+}}$) for the events $+_a$
and $-_{\dot{a}}$ ($+_b$
and $-_{\dot{b}}$). Henceforth, we will use the short hand
$\{C_i^a\}_{i\in I^a}$ ($\{C_i^b\}_{i\in I^b}$) for
$\{C_i^{a\dot{a}+-}\}_{i\in I^{a\dot{a}+-}}$
($\{C_i^{\dot{b}b-+}\}_{i\in I^{\dot{b}b-+}}$). With this notation, we get:
\begin{align}
   p_{a,\dot{a}}(+_a,-_a|C^a_i) &=
   p_{a,\dot{a}}(+_a|C^a_i)
   p_{a,\dot{a}}(-_a|C^a_i), \forall i \in I^a,  \nonumber \\
p_{\dot{b},b}(-_b,+_b|C^b_i) &=
   p_{\dot{b},b}(-_b|C^b_i)
   p_{\dot{b},b}(+_b|C^b_i), \forall i \in I^b.
\end{align}

\begin{assumption}\label{no-cons}
\begin{align}
  p(aC_i^a) &= p(a)p(C_i^a) \\
p(b C_i^b) &=p(b)p(C_i^b) \\
  p(abC_i^a) &= p(ab) p(C_i^a),  \\
  p(abC_i^b) &= p(ab) p(C_i^b), \\
  p(ab C_i^a C_j^b) &= p(ab)p(C_i^a C_j^b).
\end{align}
\end{assumption}
With this assumption one would like to exclude that the common causes are
causally relevant for the setting of the measurement apparatuses or
vice versa. Furthermore, one would like to exclude a common cause for
these factors.

\subsection{Constraints for $p_{a,b}(+_a+_b)$, $p(+_a|a)$, and
  $p(+_b|b)$}\label{constraints}
To obtain a Bell-type inequality we need an upper and a lower bound for
\begin{equation}
  p_{a,b}(+_a+_b), p(+_a|a), \mbox{ and } p(+_b|b).
\end{equation}
We will need the following proposition.
\begin{prop}\label{prop1}
Let two events $A$ and $B$ with $p(A)=p(B)=0.5$ be almost perfectly
correlated ($p(A|B)=1-\epsilon$) and assume a common cause 
$C=\{C_i\}_{i\in I}$, such that
\begin{align}\label{16}
p(AB|C_i)=p(A|C_i )p(B|C_i), \forall i\in I.
\end{align}
Then
\begin{align}\label{16aa}
\sum_{i\in I_{1}}p(C_i)-\sqrt{\epsilon}\le p(A)<\sum_{i\in I_{1}}p(C_i)+4\sqrt{\epsilon}-2\epsilon,
\end{align}
where
\begin{align}
I_{1}&:=\{i\in I: p(A|C_i)\ge 1-\sqrt{\epsilon}\}.
\end{align}
\end{prop}
We prove proposition \ref{prop1} in appendix \ref{appendix}.

With the definition 
\begin{equation}
  C:=\vee_{i\in I_1}C_i,
\end{equation}
equation~(\ref{16aa}) reads
\begin{equation}
  p(C) - \sqrt{\epsilon} \leq p(A) < p(C) +
  4\sqrt{\epsilon}-2\epsilon,
\end{equation}
or, equivalently,
\begin{equation}
  p(A) - 4\sqrt{\epsilon}+2\epsilon  < p(C) \leq p(A) +
 \sqrt{\epsilon}.
\end{equation}
We define
\begin{align}\label{def28}
I_1^a&:=\{i\in
I^a :p_{a,\dot{a}}(+_a|C_i^a)\stackrel{(*)}{=}p(+_a|aC_i^a)\geq
1-\sqrt{\epsilon_a}\}, \nonumber \\ 
C^a&:=\vee_{i\in I_1^a}C_i^a, \nonumber \\
I_1^b&:=\{i\in I^b:p_{\dot{b},b}(+_b|b C_i^b)\stackrel{(*)}{=}p(+_b|bC_i^b)\geq
1-\sqrt{\epsilon_b}\}, \nonumber \\
C^b&:=\vee_{i\in I_1^b}C_i^b.
\end{align}
In $(*)$, we use LOC.

With the substitutions
\begin{alignat}{3}
p(\dots)&\rightarrow p_{a,\dot{a}}(\dots), &\qquad p(\dots)&\rightarrow p_{\dot{b},b}(\dots), \nonumber \\
A&\rightarrow +_a, &\quad A&\rightarrow +_b, \nonumber \\
B&\rightarrow -_{\dot{a}}, &\quad B&\rightarrow -_{\dot{b}}, \nonumber
\\
C&\rightarrow C^a,&\quad C&\rightarrow C^b,
\nonumber \\
\epsilon &\rightarrow \epsilon_a, &\quad \epsilon &\rightarrow \epsilon_b,
\end{alignat}
we get
\begin{alignat}{2}
p_{a,\dot{a}}(+_a)-4\sqrt{\epsilon_a}+2\epsilon_a &< p_{a,\dot{a}}(C^a)
&\leq p_{a,\dot{a}}(+_a)+\sqrt{\epsilon_a}, \nonumber \\
p_{\dot{b},b}(+_b)-4\sqrt{\epsilon_b}+2\epsilon_b &< p_{\dot{b},b}(C^b)
&\leq p_{\dot{b},b}(+_b)+\sqrt{\epsilon_b}.
\end{alignat}
Using assumption \ref{loc2}, \ref{no-cons}, and the definition 
\begin{align}
\epsilon :=\max_{a,b} \{\epsilon_a,\epsilon_b\},
\end{align}
we get
\begin{alignat}{2}\label{38a}
p(+_a|a)-\Delta^+ &< p(C^a)
&\leq p(+_a|a)+\Delta^- , \nonumber \\
p(+_b|b)-\Delta^+ &< p(C^b)
&\leq p(+_b|b)+\Delta^-,
\end{alignat}
with
\begin{align}\label{39a}
\Delta^+&=4\sqrt{\epsilon}-2\epsilon, \nonumber \\
\Delta^-&=\sqrt{\epsilon}.
\end{align}

Using again assumptions \ref{loc2}
and \ref{no-cons}, the following bounds for $p(+_a
+_b|ab)$ can be derived (this is shown in appendix \ref{bounds}):

\begin{align}\label{39a}
p_{a,b}(+_a+_b)-\Delta_{a,b}^{+}< p(C^a C^b)\le p_{a,b}(+_a+_b)+\Delta_{a,b}^{-},
\end{align}
with
\begin{align}\label{90}
\Delta_{a,b}^{+}&=\frac{(p(a)+p(b))(5\sqrt{\epsilon}-2\epsilon)}{p(ab)},
\nonumber \\
\Delta_{a,b}^{-}&=\frac{(p(a)+p(b))\sqrt{\epsilon}}{p(ab)}.
\end{align}

\subsection{A weak Clauser-Horne inequality}\label{inequality}
In the next step, we make use of a constraint, which holds for the
probabilities of arbitrary events. For events $A$ and $B$ to be elements of
a classical probability space, it is not enough that
\begin{align}
0\leq &p(A)\leq 1, \nonumber \\
0\leq &p(B)\leq 1, \nonumber \\
0\leq &p(AB)\leq 1,
\end{align}
and
\begin{align}
p(AB)&\leq p(A), \nonumber \\
p(AB)&\leq p(B).
\end{align}
We note first, that (``$\bar{A}$'' means ``not $A$'')
\begin{align}\label{78}
p(AB)+p(A\bar{B})+p(\bar{A}B)+p(\bar{A}\bar{B})=1.
\end{align}
It is also
\begin{align}
p(A)&=p(AB)+p(A\bar{B}), \quad \mbox{and} \nonumber \\
p(B)&=p(AB)+p(\bar{A}B),
\end{align}
and hence
\begin{align}
p(A)+p(B)-p(AB)=p(AB)+p(A\bar{B})+p(\bar{A}B),
\end{align}
which implies with equation (\ref{78})\footnote{This can also be seen by noting that
  $p(A)+p(B)-p(AB)$ is the probability of the disjunction of $A$ and
  $B$, $p(A\vee B)$.}
\begin{align}
0\le p(A)+p(B)-p(AB)\le 1.
\end{align}
For more than two events there are more constraints in the form
of such inequalities.\footnote{For a detailed discussion and
  the beautiful connection to the geometry of convex polytopes, see
  e.g.~\citeA{pitowsky89}.} For four events $A$, $A'$, $B$ and $B'$,
one constraint reads
\begin{align}\label{ch}
-1\le p(AB)+p(AB')+p(A'B')-p(A'B)-p(A)-p(B')\le 0.
\end{align}
This is the Clauser-Horne inequality\footnote{What \citeA{clauser74}
  have actually derived is inequality (\ref{100}) without the
  correction terms (the $\Delta$s).  In (\ref{100}), there are
  conditional probabilities involved. Nevertheless, we adopt common terminology and
refer to both inequalities with the same name, since it will always be clear
from the context which is meant.} \cite{clauser74}, which we prove
in appendix \ref{appch}. This
inequality is an a priori constraint for arbitrary events. Hence,
for the measurement directions $1,2\in D_A$ and $3,4\in D_B$, it is also
\begin{align}
-1 \le &
p(C^1 C^3)+p(C^1 C^4)+
p(C^2 C^4)
\nonumber \\
&-p(C^2 C^3)-
p(C^1)-p(C^4) \nonumber\\ \le& 0.
\end{align}
Together with inequality (\ref{39a})
\begin{align}
p_{a,b}(+_a+_b)-\Delta_{a,b}^{+}<
p(C^a C^b) \le p_{a,b}(+_a+_b) +\Delta_{a,b}^{-} \nonumber
\end{align}
and inequality (\ref{38a})
\begin{alignat}{2}
p(+_a|a)-\Delta^{+}&< p(C^a)
&\le p(+_a|a)+\Delta^{-}, \nonumber \\
p(+_b|b)-\Delta^{+}&< p(C^b)
&\le p(+_b|b)+\Delta^{-} \nonumber
\end{alignat}
one gets
\begin{align}\label{100}
-1-\Delta_{1,3}^{-}&-\Delta_{1,4}^{-}-\Delta_{2,4}^{-}-\Delta_{2,3}^{+}-2\Delta^{+}
\nonumber \\
<& p_{1,3}(+_1+_3)
+p_{1,4}(+_1+_4)+p_{2,4}(+_2+_4)
\nonumber \\
&-p_{2,3}(+_2+_3)-p(+_1|1)-p(+_4|4)
\nonumber \\
<&
\Delta_{1,3}^{+}+\Delta_{1,4}^{+}+\Delta_{2,4}^{+}+\Delta_{2,3}^{-}+2\Delta^{-}, 
\end{align}
with 
\begin{align}
\Delta_{a,b}^{-}&=\frac{(p(a)+p(b))\sqrt{\epsilon}}{p(ab)}, \nonumber \\
\Delta_{a,b}^{+}&=\frac{(p(a)+p(b))(5\sqrt{\epsilon}-2\epsilon)}{p(ab)}, \nonumber \\
\Delta^{-}&=\sqrt{\epsilon}, \nonumber \\
\Delta^{+}&=4\sqrt{\epsilon}-2\epsilon .
\end{align}
Note that this inequality reduces to the Clauser-Horne inequality
for $\epsilon =0$.

\subsection{Contradiction}\label{violation}

The predicted values of $p_{1,3}(+_1,+_3)$, $p_{1,4}(+_1,+_4)$,
$p_{2,4}(+_2,+_4)$ and $p_{2,3}(+_2,+_3)$ by QM are such that the
maximal violation\footnote{These violations are also maximal in that
  no other quantum mechanical two-particle state for two
  spin-$\frac{1}{2}$-particles yields a bigger violation (see
  \citeA{tsirelson80} and \citeA{cabello02}).} for the lower bound of
(\ref{100}) occurs (among others) for the angles
$\varphi_{1,3}=\varphi_{1,4}=\varphi_{2,4}=\frac{\pi}{4}$ and 
$\varphi_{2,3}=\frac{3\pi}{4}$:
\begin{align}
-1-\Delta_{1,3}^{-}&-\Delta_{1,4}^{-}-\Delta_{2,4}^{-}-\Delta_{2,3}^{+}-\Delta^{+}-\Delta^{+}
\nonumber \\
< &  p_{1,3}(+_1+_3)
+p_{1,4}(+_1+_4)+p_{2,4}(+_2,+_4) \nonumber 
\nonumber \\
&-p_{2,3}(+_2+_3)-p(+_1|1)-p(+_4|4)=-\frac{\sqrt{2}+1}{2}.
\end{align} 
The maximal violation for the upper bound occurs (among others) for the angles
$\varphi_{1,3}=\varphi_{2,4}=\frac{3\pi}{4}$,
$\varphi_{1,4}=\frac{5\pi}{4}$ and $\varphi_{2,3}=\frac{\pi}{4}$:
\begin{align}
\frac{\sqrt{2}-1}{2}=&p_{1,3}(+_1+_3)
+p_{1,4}(+_1+_4)+p_{2,4}(+_2+_4)
\nonumber \\
&-p_{2,3}(+_2+_3)-p(+_1|1)-p(+_4|4)
\nonumber \\
<&
\Delta_{1,3}^{+}+\Delta_{1,4}^{+}+\Delta_{2,4}^{+}+\Delta_{2,3}^{-}+\Delta^{-}+\Delta^{-}, 
\end{align}

With $p(ab)=\frac{1}{4}$ and $p(a)=p(b)=\frac{1}{2}$, one has
\begin{align}
\Delta_{a,b}^-=4\sqrt{\epsilon} \quad \mbox{and} \quad
\Delta_{a,b}^+=20\sqrt{\epsilon}-8\epsilon .
\end{align}
With the chosen angles and measurement probabilities one gets
\begin{align}
\frac{\sqrt{2}-1}{2}< 40\sqrt{\epsilon}-12\epsilon.
\end{align}
for the
lower bound.
This inequality is violated for
\begin{align}
\epsilon \le \epsilon^l_{\max}, \quad \epsilon^l_{\max}=2.689 \cdot 10^{-5}.
\end{align}
The inequality for the upper bound reads
\begin{align}
\frac{\sqrt{2}-1}{2} < 66\sqrt{\epsilon}-24\epsilon,
\end{align}
which is violated for
\begin{align}
\epsilon \le \epsilon^u_{\max}, \quad \epsilon^u_{\max}=9.869 \cdot 10^{-6}. 
\end{align}

Thus the quantum mechanical predictions
contradict the predictions of a hidden variable model obeying our assumptions
for
\begin{align}
\epsilon \le \epsilon^l_{\max} = 2.698 \cdot 10^{-5}. 
\end{align}

\section{Discussion}\label{discussion}

\begin{figure}
\begin{center}
\epsfig{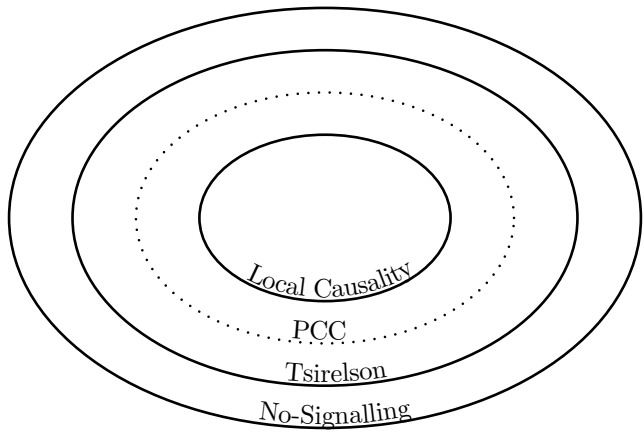}
\caption{\label{fig3}Comparison with other constraints.}
\end{center}
\end{figure}

Even though the four sets of assumptions in \citeA{bell64},
\citeA{clauser69}, \citeA{bell71} and \citeA{grasshoff05} (see figure
\ref{fig1}) differ, they all imply the same constraints on the
correlations as expressed in the Clauser-Horne
inequality.\footnote{Even though the derived inequalities have a
  different form in \citeA{bell64}, \citeA{bell71} and
  \citeA{grasshoff05}, the assumptions are all sufficient to derive
  also the Clauser-Horne inequality.}  

One of the questions left open by \citeA{grasshoff05} is what constraints are
implied \emph{without} assuming the existence of perfectly correlated
events. Since with a slightest deviation from perfect correlations the proof
by \citeA{grasshoff05} breaks down, it gives no hint as to whether the same
Bell-type constraints follow nor whether a contradiction to QM is entailed at all.
In the present paper we have given a partial answer to that
question. 

The inequality we get at the end of our derivation is stronger than the
quantum mechanical predictions (for $\epsilon \le \epsilon^l_{\max}$), but
weaker than the Clauser-Horne inequality (for $\epsilon>0$). Thus,
the weakening of the assumptions is also reflected in a resulting
weakening of the constraints. Note however that we did not prove that
this weakening is really an implication of our assumptions. What we have
shown is only that the conditional probabilities \emph{at least} have to obey the constraint (\ref{100}).\footnote{This proviso is also necessary, because some
steps can be optimized in our derivation. For example, one can choose the
  borders of the partitions in (\ref{1}) and (\ref{a82}) differently,
such that one would get tighter constraints, that is, the correction
terms to the Clauser-Horne inequality (the $\Delta$'s) would become
smaller.} To prove the stronger proposition, one could try to
construct 
a separate common cause model obeying all our assumptions that
violates the Clauser-Horne inequality without correction terms,
but does not violate the weak Clauser-Horne inequality.

Now, we would like to compare our inequality to other prominent
constraints (see figure \ref{fig3}).\footnote{For an overview, see
  e.g.~\citeA{gisin05}.} Even though Bell's
theorem excludes models which obey local causality, the predictions of QM
for $p(A_a B_b|ab)$ still obey the no-signalling constraint\footnote{See e.g. \citeA{redhead87}, pp.~113-117 and
  references therein.}
\begin{align}\label{nosig}
\sum_{B_b}p(A_a B_b|ab)&=\sum_{B_b}p(A_aB_b|ab'), \nonumber \\
\sum_{A_a}p(A_a B_b|ab)&=\sum_{A_{a'}}p(A_{a'}B_b|a'b),
\end{align}
which states that the probability of the measurement outcome on one side does not
depend on the measurement direction on the other side (given the
quantum mechanical state of the system). Moreover, there are some
correlations obeying no-signalling which are not permitted by quantum
mechanics.\footnote{By this we mean possible predictions for the
  probabilities of $p_{a,b}(A_a B_b)$ coming from Hilbert space-vectors.} The bounds, which are allowed by quantum mechanics, were first
derived by \citeA{tsirelson80}. Notoriously, still more constraining are the Bell inequalities. The situation is drawn schematically
in figure \ref{fig3}. The boarder at the margin is the least constraining
coming from the no-signalling condition (\ref{nosig}). Next is the
Tsirelson-bound, which is again weaker than the bound coming from the
Clauser-Horne inequality (local causality). The bound coming from inequality
(\ref{100}) lies between the
Tsirelson bound and the bound coming from the Clauser-Horne
inequality, depending on the value of $\epsilon$. For $\epsilon\neq 0$ there are quantum mechanical
states which do violate the Clauser-Horne inequality but not (\ref{100}). This reveals the following a priori
possibility. As \citeA{gisin91} showed, the correlations coming from
pure entangled states always violate the Clauser-Horne
inequality. For $\epsilon\neq 0$, Gisin's argument is not sufficient
to conclude that all entangled states violate inequality
(\ref{100}). Hence, it is an open question, whether or not there exist
models obeying all our assumptions, for the correlations of some
entangled pure states.

Even though $\epsilon^l_{\max}$ is not zero, it is still very
small. Particularly, violations of correlations deviating only
through $\epsilon^l_{\max}$ from being perfect, can experimentally
not be ruled out. This means that we cannot rule out the existence of an empirically adequate hidden variable
model obeying all our assumptions. On the other hand, from a
theoretical point of view, a deviation from perfect correlation of
order $\epsilon^l_{\max}$ is rather big. Modulo some theoretical
assumptions, any non-vanishing $\epsilon$ can be interpreted as a
violation of rotation invariance (see section \ref{constraints}),
which moreover induces a violation of Lorentz invariance.\footnote{Whether
or not a violation under rotation invariance implies also a violation
of Lorentz boost invariance is model dependent (see
\citeA{mattingly05}).} 
Triggered by theoretical
works in various approaches to quantum gravity, which either imply
violations of Lorentz invariance or render such a violation natural,
there has been a tremendous experimental effort for finding signatures of
such violations during the last ten years or so (for a recent review,
see e.g.~\citeA{mattingly05}). The constraints coming from negative
results of such experiments are rather strong. In view of these
findings, one would expect $\epsilon$ to be smaller than
$\epsilon^l_{\max}$ and the inequality (\ref{100}) to be violated.

\section*{Acknowledgments}
We would like to thank Nicolas Gisin, Gerd Gra\ss hoff, Stephanie
Kurmann, Peter Minkowski and the audience of the ``Workshop Philosophy of Physics'',
Lausanne, 14th of March, for discussions, G\'abor
Hofer-Szab\'o for  correspondence in connection with this article, and
Michael Baumgartner and Tim
R\"az for discussions and proofreading.

\begin{appendix}
\section{Proof of proposition \ref{prop1}}\label{appendix}
\begin{prop1}
Let two events $A$ and $B$ with $p(A)=p(B)=0.5$ be almost perfectly
correlated ($p(A|B)=1-\epsilon$) and assume a common cause
$C=\{C_i\}_{i\in I}$, such that
\begin{align}\label{n65}
p(AB|C_i )=p(A|C_i )p(B|C_i ),\quad  \forall i\in I.
\end{align}
Then
\begin{align}
\sum_{i\in I_{1}}p(C_i)-\sqrt{\epsilon}\le p(A)<\sum_{i\in
  I_{1}}p(C_i)+4\sqrt{\epsilon}-2\epsilon ,
\end{align}
where
\begin{align}
I_{1}&:=\{i\in I: p(A|C_i)\ge 1-\sqrt{\epsilon}\}.
\end{align}
\end{prop1}
\begin{proof}
We will partition $I$ into the
following
three subsets:
\begin{align}\label{1}
I_{1}&:=\{i\in I: p(A|C_i)\ge 1-\sqrt{\epsilon}\}, \nonumber \\
I_{2}&:=\{i\in I: \sqrt{\epsilon}<p(A|C_i)< 1-\sqrt{\epsilon}\},
\nonumber \\
I_{3}&:=\{i\in I: p(A|C_i)\le \sqrt{\epsilon}\}.
\end{align}
It is
\begin{align}
p(A)=\sum_{i\in I_{1}} p(A|C_i)p(C_i)+\sum_{i\in I_{2}} p(A|C_i)p(C_i)+\sum_{i\in I_{3}} p(A|C_i)p(C_i).
\end{align}
With the definitions (\ref{1}) the following inequalities hold:
\begin{align}
p(A)&\ge \sum_{i\in I_{1}}p(A|C_i)p(C_i)\ge
(1-\sqrt{\epsilon})\sum_{i\in I_{1}}p(C_i)\ge \sum_{i\in
  I_{1}}p(C_i)-\sqrt{\epsilon}, \nonumber \\
p(A)&\le \sum_{i\in I_{1}}p(C_i)+\sum_{i\in I_{2}}p(A|C_i)p(C_i)+\sqrt{\epsilon}.
\end{align}
Hence, to complete the proof we have to show that
\begin{align}
\sum_{i\in I_{2}}p(A|C_i)p(C_i)< 3\sqrt{\epsilon}-2 \epsilon.
\end{align}
It is
\begin{align}
\frac{\epsilon}{2}=\underbrace{p(A)}_{\frac{1}{2}}-\underbrace{p(AB)}_{\frac{1-\epsilon}{2}}
& =\sum_{i\in I} \left[ p(A|C_i)-p(AB|C_i)\right] p(C_i) \nonumber \\
&\stackrel{(*)}{=}\sum_{i\in I} \left[ p(A|C_i)-p(A|C_i)p(B|C_i)\right]
p(C_i) \nonumber \\
& =\sum_{i\in I}p(A|C_i)\left[ 1-p(B|C_i)\right] p(C_i),
\end{align}
where we used (\ref{n65}) to get equality $(*)$. Since everything is
symmetric in $A$ and $B$, the same holds if one exchanges $A$ and $B$
for each other. We thus have
\begin{align}
\sum_{i\in
  I}p(A|C_i)\left[1-p(B|C_i)\right]p(C_i)&=\frac{\epsilon}{2},
\label{2.1} \\
\sum_{i\in
  I}p(B|C_i)\left[1-p(A|C_i)\right]p(C_i)&=\frac{\epsilon}{2}. \label{2.2}
\end{align}
Since all terms in the sums on the L.H.S.~of eq.~(\ref{2.1}, \ref{2.2}) are
positive, the following inequalities hold for all subsets $I^{\subset}$
of the value space $I$:
\begin{align}
0\le \sum_{i\in
   I^{\subset}}p(A|C_i)\left[1-p(B|C_i)\right]p(C_i)&\le\frac{\epsilon}{2}, \quad \forall I^{\subset}\subset I, \label{3}\\
0\le \sum_{i\in
  I^{\subset}}p(B|C_i)\left[1-p(A|C_i)\right]p(C_i)&\le\frac{\epsilon}{2}, \quad \forall I^{\subset}\subset I.
\label{4}
\end{align}
Subtracting (\ref{3}) from (\ref{4}), one gets
\begin{align}\label{5}
\left| \sum_{i\in I^{\subset}}\left[
    p(A|C_i)-p(B|C_i)\right]p(C_i)\right|\le \frac{\epsilon}{2},
\quad \forall I^{\subset}\subset I.
\end{align}

With the definitions
\begin{align}
I_{2}^{A\ge B}&:=\left\{ i\in I_{2}:p(A|C_i)\ge p(B|C_i)\right\}, \\
I_{2}^{A< B}&:=\left\{ i\in I_{2}:p(A|C_i)< p(B|C_i)\right\}
\end{align}
and applying (\ref{5}) for these sets, one gets

\begin{align}
\left| \sum_{i\in I_{2}^{A\ge B}}\left[
    p(A|C_i)-p(B|C_i)\right] p(C_i)\right|&=  \sum_{i\in I_{2}^{A\ge B}}\left|
    p(A|C_i)-p(B|C_i)\right|p(C_i)\le \frac{\epsilon}{2}, \\
\left| \sum_{i\in I_{2}^{A< B}}\left[
    p(A|C_i)-p(B|C_i)\right]p(C_i)\right|&=  \sum_{i\in I_{2}^{A< B}}\left|
    p(A|C_i)-p(B|C_i)\right|p(C_i)\le \frac{\epsilon}{2}.
\end{align}
Adding these two inequalities, one gets:
\begin{align}\label{6}
\sum_{i\in I_2
}\left|
    p(A|C_i)-p(B|C_i)\right|p(C_i)\le \epsilon .
\end{align}
We partition $I_{2}$ in the following two subsets:
\begin{align}\label{a82}
I_{2}^{\ge\sqrt{\epsilon}}&:=\left\{ i\in I_{2}: \left|p(A|C_i)-p(B|C_i)
\right|\ge \frac{\sqrt{\epsilon}}{2}\right\}, \nonumber \\
I_{2}^{<\sqrt{\epsilon}}&:=\left\{ i\in I_{2}: \left|p(A|C_i)-p(B|C_i)
\right|< \frac{\sqrt{\epsilon}}{2}\right\}.
\end{align}
From
\begin{align}
\sum_{i\in I_{2}^{\ge\sqrt{\epsilon}}}\left|p(A|C_i)-p(B|C_i)
\right| p(C_i)\ge \frac{\sqrt{\epsilon}}{2}\sum_{i\in I_{2}^{\ge\sqrt{\epsilon}}}p(C_i)
\end{align}
together with (\ref{6}), we get
\begin{align}\label{7}
\sum_{i\in  I_{2}^{\ge\sqrt{\epsilon}}}p(C_i)\le 2\sqrt{\epsilon}.
\end{align}
Remember, that we want to derive an upper bound for
\begin{align}
\sum_{i\in I_{2}}p(A|C_i)p(C_i).
\end{align}
With (\ref{7}), we already have
\begin{align}\label{7.2}
\sum_{i\in I_{2}}p(A|C_i)p(C_i)&=\sum_{i\in
   I_{2}^{\ge\sqrt{\epsilon}}}p(A|C_i)p(C_i)+\sum_{i\in
   I_{2}^{<\sqrt{\epsilon}}}p(A|C_i)p(C_i) \nonumber \\
&< (1-\sqrt{\epsilon})2\sqrt{\epsilon}+\sum_{i\in
   I_{2}^{<\sqrt{\epsilon}}}p(A|C_i)p(C_i).
\end{align}
We will use again inequality (\ref{3}), this time for the set $I_{2}^{<\sqrt{\epsilon}}$:
\begin{align}\label{8}
\sum_{i\in  I_{2}^{<\sqrt{\epsilon}}
   }p(A|C_i)\left[1-p(B|C_i)\right]p(C_i)&\le\frac{\epsilon}{2}.
\end{align}
Because we are looking at the subset $I_{2}^
{<\sqrt{\epsilon}}$, it is
\begin{align}
\sum_{i\in  I_{2}^{<\sqrt{\epsilon}}
   }p(A|C_i)\left[1-p(B|C_i)\right]p(C_i)> \sum_{i\in  I_{2}^{<\sqrt{\epsilon}}
   }p(A|C_i)\left[1-p(A|C_i)-\frac{\sqrt{\epsilon}}{2}\right]p(C_i).
\end{align}
With (\ref{8}), one gets
\begin{align}\label{113}
\sum_{i\in
  I_{2}^{<\sqrt{\epsilon}}}p(A|C_i)\left[1-p(A|C_i)-\frac{\sqrt{\epsilon}}{2}\right]p(C_i)&<\frac{\epsilon}{2}.
\end{align}

Now, since $I_{2}^{<\sqrt{\epsilon}}$ is a subset of $I_{2}$,
$p(A|C_i)$ takes on values in the interval $[ \sqrt{\epsilon},
1-\sqrt{\epsilon}]$. One can check that each summand is certainly greater than for
$p(A|C_i)=1-\sqrt{\epsilon}$. We have
\begin{align}
\sum_{i\in
  I_{2}^{<\sqrt{\epsilon}}}p(A|C_i)\left[1-p(A|C_i)-\frac{\sqrt{\epsilon}}{2}\right]p(C_i)>(1-\sqrt{\epsilon})\frac{\sqrt{\epsilon}}{2}\sum_{i\in
  I_{2}^{<\sqrt{\epsilon}}}p(C_i).
\end{align}
We get the constraint
\begin{align}
\sum_{i\in
  I_{2}^{<\sqrt{\epsilon}}}p(C_i)<\frac{\sqrt{\epsilon}}{
  (1-\sqrt{\epsilon})}.
\end{align}
With (\ref{7.2}) one gets
\begin{align}
\sum_{i\in
  I_{2}}p(A|C_i)p(C_i)<(1-\sqrt{\epsilon})2\sqrt{\epsilon}+\sqrt{\epsilon}=3\sqrt{\epsilon}-2\epsilon ,
\end{align}
which is what we wanted to show. We have
\begin{align}\label{39}
\sum_{i\in I_{1}}p(C_i)-\sqrt{\epsilon}\le p(A)<\sum_{i\in
  I_{1}}p(C_i)+4\sqrt{\epsilon}-2\epsilon .
\end{align}
\end{proof}

\section{Bounds for $p(+_a+_b|ab)$}\label{bounds}
With (\ref{38a}) and assumption \ref{no-cons}, we get
\begin{align}
p(+_a|a)&=p(+_aC^a|a)+p(+_a\overline{C^a}|a)<
p(C^a|a) +4\sqrt{\epsilon}-2\epsilon   \nonumber \\
&=p(+_a C^a|a)+ p(\overline{+_a} C^a|a)+4\sqrt{\epsilon}-2\epsilon, 
\end{align}
and hence
\begin{align}\label{92}
p(+_a a\overline{C^a})< p(\overline{+_a}a
C^a)+p(a)\left( 4\sqrt{\epsilon}-2\epsilon \right).
\end{align}
Furthermore, we have

\begin{align}\label{93}
p(\overline{+_a} C^a|a)&\stackrel{(*)}{=}\sum_{i\in
  I_1^a}p(\overline{+_a} |a C^a_i)p(C^a_i) \nonumber \\
&=\sum_{i\in
  I_1^a}\left( 1-p(+_a |a C^a_i)\right) p(C^a_i)\nonumber \\ 
&\leq \sqrt{\epsilon}\sum_{i\in
  I_1^a}p(C^a_i)\leq \sqrt{\epsilon},
\end{align}
where we used assumption \ref{no-cons} to get equality $(*)$. We can
write (\ref{93}) as

\begin{align}\label{94}
p(\overline{+_a} a C^a)\leq p(a)\sqrt{\epsilon},
\end{align}
such that we get from (\ref{92})

\begin{align}\label{95}
p(+_a a\overline{C^a} X)< p(a)\left( 5\sqrt{\epsilon}-2\epsilon \right),
\end{align}
because for any $X$ and any $Y$, $p(XY)\leq p(Y)$. Next, from

\begin{align}
p(+_aa C^a  X)=p(a C^a  X)-p(\overline{+_a}a C^a  X)
\end{align}
together with (\ref{94}) and because $p(\overline{+_a}a C^a
X)\leq p(\overline{+_a}a C^a  )$ we get
\begin{align}\label{97}
p(+_a a C^a X)\geq p(a C^a  X)-p(a) \sqrt{\epsilon}.
\end{align}
Starting from the other inequalities in (\ref{38a}), we get
inequalities analogue to (\ref{95}) and (\ref{97}). We have
\begin{align}
p(+_a a \overline{C^a} X)&< p(a)\left(
  5\sqrt{\epsilon}-2\epsilon \right), \label{98a}\\
p(+_b b \overline{C^b} X)&< p(b)\left(
  5\sqrt{\epsilon}-2\epsilon \right), \label{99a}\\
p(+_a a C^a  X)&\geq p(a C^a  X)-p(a) \sqrt{\epsilon}, \label{100a}\\
p(+_b b C^b  X)&\geq p(b C^b  X)-p(b) \sqrt{\epsilon}. \label{101a}
\end{align}

Now, we can derive an upper bound for $p\left( +_a+_bab\right)$, using
(\ref{98a}) and (\ref{99a}): 
\begin{equation}
  \begin{split}
    p(+_a+_bab) &= p(+_a+_b abC^a) +
    p(+_a+_b ab\overline{C^a}) \\
    &< p(+_a+_b abC^a) + p(a)\left(5\sqrt{\epsilon}-2\epsilon\right)  \\
    &< p(+_a+_babC^a C^b) + \left(p(a)+p(b)\right)
    \left(5\sqrt{\epsilon}-2\epsilon\right) \\
    &\leq p(abC^aC^b) +
    \left(p(a)+p(b)\right)
    \left(5\sqrt{\epsilon}-2\epsilon\right) .
  \end{split}
\end{equation}

Using also assumption \ref{no-cons}, we finally obtain
   \begin{eqnarray}\label{106a}
       p(+_a+_b|ab)
       &\equiv & \frac{p(+_a +_b ab)}{p(ab)} \nonumber \\
     &<& \frac{p(ab C^a
        C^b) +
       \left(p(a)+p(b)\right)(5\sqrt{\epsilon}-2\epsilon)}{p(ab)} \nonumber \\
       &= & p(C^a C^b|ab) + \left(p(a)+p(b)\right)\frac{5\sqrt{\epsilon}-2\epsilon}{p(ab)} \nonumber \\ 
       &=& p(C^aC^b) + \left(p(a)+p(b)\right)\frac{5\sqrt{\epsilon}-2\epsilon}{p(ab)}.  
   \end{eqnarray}
Next, we derive a lower bound.

\begin{equation}
  \begin{split}
    p(+_a+_bab) &\geq
    p(+_a+_babC^aC^b)\\ 
    &\geq  p(+_babC^aC^b) - p(a)\sqrt{\epsilon} \\
    &\geq p(abC^aC^b) - \left(p(a)+p(b)\right)\sqrt{\epsilon},
  \end{split}
\end{equation}

   \begin{eqnarray}\label{109a}
       p(+_a+_b|ab)
       &\equiv & \frac{p(+_a +_b ab)}{p(ab)} \nonumber \\
     &\geq& p(C^aC^b) - \frac{\left(p(a)+p(b)\right)\sqrt{\epsilon}}{p(ab)}. 
   \end{eqnarray}

(\ref{106a}) and (\ref{109a}) imply
\begin{align}\label{89}
p(+_a+_b|ab)-\Delta_{a,b}^{+}< p(C^aC^b)\le p(+_a+_b|ab)+\Delta_{a,b}^{-},
\end{align}
with
\begin{align}\label{90}
\Delta_{a,b}^{-}&=\frac{\left(p(a)+p(b)\right)\sqrt{\epsilon}}{p(ab)}, \nonumber \\
\Delta_{a,b}^{+}&=\frac{\left(p(a)+p(b)\right)(5\sqrt{\epsilon}-2\epsilon)}{p(ab)}.
\end{align}

\section{Proof of the Clauser-Horne Inequality}\label{appch}
In this appendix we will prove inequality (\ref{ch}). 
We consider arbitrary four events $A$, $A'$, $B$, and $B'$
together with their complements. The sum over all $16$ possibilities
equals one:
\begin{align}\label{sum}
\sum_{a,a',b,b'}p(a,a',b,b')=1, \qquad \mbox{where} \quad a\in \{A,
\bar{A}\} \quad \mbox{etc.}
\end{align}
We also have
\begin{align}
p(AB)=&p(AA'BB')+p(AA'B\bar{B'})+p(A\bar{A'}BB')+p(A\bar{A'}B\bar{B'}),
\nonumber \\
p(AB')=&p(AA'BB')+p(A\bar{A'}BB')+p(AA'\bar{B}B')+p(A\bar{A'}\bar{B}B'),
\nonumber \\
p(A'B')=&p(AA'BB')+p(AA'\bar{B}B')+p(\bar{A}A'BB')+p(\bar{A}A'\bar{B}B'), 
\nonumber \\
p(A'B)=&p(AA'BB')+p(AA'B\bar{B'})+p(\bar{A}A'BB')+p(\bar{A}A'B\bar{B'}),
\nonumber \\
p(A)=&p(AA'BB')+p(AA'B\bar{B'})+p(A\bar{A'}BB')+p(A\bar{A'}B\bar{B'})
\nonumber \\
&+p(AA'\bar{B}B')+p(AA'\bar{B}\bar{B'})+p(A\bar{A'}\bar{B}B')+p(A\bar{A'}\bar{B}\bar{B'}),
\nonumber \\
p(B')=&p(AA'BB')+p(A\bar{A'}BB')+p(AA'\bar{B}B')+p(A\bar{A'}\bar{B}B')
\nonumber \\
&+p(\bar{A}A'BB')+p(\bar{A}\bar{A'}BB')+p(\bar{A}A'\bar{B}B')+p(\bar{A}\bar{A'}\bar{B}B').
\end{align}
Thus
\begin{align}\label{120}
p(AB)+&p(AB')+p(A'B')-p(A'B)-p(A)-p(B')  \nonumber \\
=&-\big[p(AA'B\bar{B'})+p(AA'\bar{B}\bar{B'})+p(A\bar{A'}\bar{B}B')+p(A\bar{A'}\bar{B}\bar{B'})
\nonumber \\
&+p(\bar{A}A'BB')+p(\bar{A}A'B\bar{B'})+p(\bar{A}\bar{A'}BB')+p(\bar{A}\bar{A'}\bar{B}B')\big].
\end{align}
Because each term appears only once on
the R.H.S.~of equation (\ref{120}), equation (\ref{sum}) implies the Clauser-Horne inequality:
\begin{align}
-1\le p(AB)+p(AB')+p(A'B')-p(A'B)-p(A)-p(B')\le 0.
\end{align}

\end{appendix}

\bibliographystyle{apacite}
\bibliography{phys-rev}

\begin{thebibliography}{}

\bibitem[\protect\citeauthoryear{%
Bell%
}{%
Bell%
}{%
{\protect\APACyear{1964}}%
}]{%
bell64}%
\APACinsertmetastar{%
bell64}%
Bell, J.~S.%
%
\newblock{}\BBOP{}1964\BBCP{}.
\newblock{}\BBOQ{}{On the Einstein-Podolsky-Rosen Paradox}.\BBCQ{}
\newblock{}\Bem{Physics}, \Bem{1}, 195.
\newblock{}(Reprinted in \cite[pp.~14-21]{bell87})

\bibitem[\protect\citeauthoryear{%
Bell%
}{%
Bell%
}{%
{\protect\APACyear{1971}}%
}]{%
bell71}%
\APACinsertmetastar{%
bell71}%
Bell, J.~S.%
%
\newblock{}\BBOP{}1971\BBCP{}.
\newblock{}\BBOQ{}{Introduction to the Hidden-Variable Question}.\BBCQ{}
\newblock{}\BIn{} \Bem{Foundations of quantum mechanics}\ (\BPG\ 171).
\newblock{}New York: Academic.
\newblock{}(Reprinted in \cite[pp.~29-39]{bell87}.)

\bibitem[\protect\citeauthoryear{%
Bell%
}{%
Bell%
}{%
{\protect\APACyear{1975}}%
}]{%
bell76}%
\APACinsertmetastar{%
bell76}%
Bell, J.~S.%
%
\newblock{}\BBOP{}1975, July 28\BBCP{}.
\newblock{}\Bem{{The theory of local beables}.}
\newblock{}TH-2053-CERN.
\newblock{}(Presented at the Sixth GIFT Seminar, Jaca, 2--7 June 1975,
  reproduced in \emph{Epistemological Letters}, March 1976, and reprinted in
  \cite[pp.~52-62]{bell87})

\bibitem[\protect\citeauthoryear{%
Bell%
}{%
Bell%
}{%
{\protect\APACyear{1987}}%
}]{%
bell87}%
\APACinsertmetastar{%
bell87}%
Bell, J.~S.%
%
\newblock{}\BBOP{}1987\BBCP{}.
\newblock{}\Bem{Speakable and unspeakable in quantum mechanics}.
\newblock{}Cambridge: Cambridge University Press.

\bibitem[\protect\citeauthoryear{%
Belnap%
\ \BBA{} Szab\'o%
}{%
Belnap%
\ \BBA{} Szab\'o%
}{%
{\protect\APACyear{1996}}%
}]{%
belnap96}%
\APACinsertmetastar{%
belnap96}%
Belnap, N.%
\BCBT{}\ \BBA{} Szab\'o, L.%
%
\newblock{}\BBOP{}1996\BBCP{}.
\newblock{}\BBOQ{}{Branching space-time analysis of the GHZ theorem}.\BBCQ{}
\newblock{}\Bem{Foundations of Physics}, \Bem{26}, 989-1002.

\bibitem[\protect\citeauthoryear{%
Bohm%
}{%
Bohm%
}{%
{\protect\APACyear{1951}}%
}]{%
bohm51}%
\APACinsertmetastar{%
bohm51}%
Bohm, D.%
%
\newblock{}\BBOP{}1951\BBCP{}.
\newblock{}\Bem{Quantum theory}.
\newblock{}New York: Prentice Hall.

\bibitem[\protect\citeauthoryear{%
Butterfield%
}{%
Butterfield%
}{%
{\protect\APACyear{1989}}%
}]{%
butterfield89}%
\APACinsertmetastar{%
butterfield89}%
Butterfield, J.%
%
\newblock{}\BBOP{}1989\BBCP{}.
\newblock{}\BBOQ{}{A Space-Time Approach to the Bell inequality}.\BBCQ{}
\newblock{}\BIn{} J.~T. Cushing\ \BBA{} E.~McMullin\ (\BEDS),
  \Bem{Philosophical consequences of quantum theory}\ (\BPG\ 114-144).
\newblock{}Notre Dame: University of Notre Dame Press.

\bibitem[\protect\citeauthoryear{%
Butterfield%
}{%
Butterfield%
}{%
{\protect\APACyear{1992}}%
{\protect\APACexlab{{\protect\BCnt{1}}}}}]{%
butterfield92}%
\APACinsertmetastar{%
butterfield92}%
Butterfield, J.%
%
\newblock{}\BBOP{}1992{\protect\BCnt{1}}\BBCP{}.
\newblock{}\BBOQ{}{Bell's theorem: What it takes}.\BBCQ{}
\newblock{}\Bem{The British Journal for the Philosophy of Science},
  \Bem{43}(1), 41-83.

\bibitem[\protect\citeauthoryear{%
Butterfield%
}{%
Butterfield%
}{%
{\protect\APACyear{1992}}%
{\protect\APACexlab{{\protect\BCnt{2}}}}}]{%
butterfield922}%
\APACinsertmetastar{%
butterfield922}%
Butterfield, J.%
%
\newblock{}\BBOP{}1992{\protect\BCnt{2}}\BBCP{}.
\newblock{}\BBOQ{}{David Lewis Meets John Bell}.\BBCQ{}
\newblock{}\Bem{Philosophy of Science}, \Bem{59}(1), 26-43.

\bibitem[\protect\citeauthoryear{%
Cabello%
}{%
Cabello%
}{%
{\protect\APACyear{2002}}%
}]{%
cabello02}%
\APACinsertmetastar{%
cabello02}%
Cabello, A.%
%
\newblock{}\BBOP{}2002\BBCP{}.
\newblock{}\BBOQ{}{Violating Bell's Inequality Beyond Cirel'son's
  Bound}.\BBCQ{}
\newblock{}\Bem{Phys. Rev. Lett.}, \Bem{88}, 060403.

\bibitem[\protect\citeauthoryear{%
Cartwright%
}{%
Cartwright%
}{%
{\protect\APACyear{1979}}%
}]{%
cartwright79}%
\APACinsertmetastar{%
cartwright79}%
Cartwright, N.%
%
\newblock{}\BBOP{}1979\BBCP{}.
\newblock{}\BBOQ{}{Causal Laws and Effective Strategies}.\BBCQ{}
\newblock{}\Bem{No\^us}, \Bem{13}, 419--437.

\bibitem[\protect\citeauthoryear{%
{Clauser}%
\ \BBA{} {Horne}%
}{%
{Clauser}%
\ \BBA{} {Horne}%
}{%
{\protect\APACyear{1974}}%
}]{%
clauser74}%
\APACinsertmetastar{%
clauser74}%
{Clauser}, J.%
\BCBT{}\ \BBA{} {Horne}, M.%
%
\newblock{}\BBOP{}1974\BBCP{}.
\newblock{}\BBOQ{}Experimental consequences of objective local theories.\BBCQ{}
\newblock{}\Bem{Physical Review D}, \Bem{10}, 526-535.

\bibitem[\protect\citeauthoryear{%
{Clauser}%
, {Horne}%
, {Shimony}%
\BCBL{}\ \BBA{} {Holt}%
}{%
{Clauser}%
\ \protect\BOthers{.}}{%
{\protect\APACyear{1969}}%
}]{%
clauser69}%
\APACinsertmetastar{%
clauser69}%
{Clauser}, J.%
, {Horne}, M.%
, {Shimony}, A.%
\BCBL{}\ \BBA{} {Holt}, R.%
%
\newblock{}\BBOP{}1969\BBCP{}.
\newblock{}\BBOQ{}{Proposed Experiment to Test Local Hidden-Variable
  Theories}.\BBCQ{}
\newblock{}\Bem{Physical Review Letters}, \Bem{23}, 880-884.

\bibitem[\protect\citeauthoryear{%
Clauser%
\ \BBA{} Shimony%
}{%
Clauser%
\ \BBA{} Shimony%
}{%
{\protect\APACyear{1978}}%
}]{%
clauser78}%
\APACinsertmetastar{%
clauser78}%
Clauser, J.%
\BCBT{}\ \BBA{} Shimony, A.%
%
\newblock{}\BBOP{}1978\BBCP{}.
\newblock{}\BBOQ{}Bell's theorem: experimental tests and implications.\BBCQ{}
\newblock{}\Bem{Rep. Prog. Phys.}, \Bem{78}, 1881-1927.

\bibitem[\protect\citeauthoryear{%
Clifton%
\ \BBA{} Kent%
}{%
Clifton%
\ \BBA{} Kent%
}{%
{\protect\APACyear{2000}}%
}]{%
clifton00}%
\APACinsertmetastar{%
clifton00}%
Clifton, R.%
\BCBT{}\ \BBA{} Kent, A.%
%
\newblock{}\BBOP{}2000\BBCP{}.
\newblock{}\BBOQ{}{Simulating Quantum Mechanics by Non-Contextual Hidden
  Variables}.\BBCQ{}
\newblock{}\Bem{Proc. Roy. Soc. Lond. A}, \Bem{456}, 2101-2114.
\newblock{}(URL = http://xxx.lanl.gov/abs/quant-ph/9908031)

\bibitem[\protect\citeauthoryear{%
Eells%
\ \BBA{} Sober%
}{%
Eells%
\ \BBA{} Sober%
}{%
{\protect\APACyear{1983}}%
}]{%
eells83}%
\APACinsertmetastar{%
eells83}%
Eells, E.%
\BCBT{}\ \BBA{} Sober, E.%
%
\newblock{}\BBOP{}1983\BBCP{}.
\newblock{}\BBOQ{}{Probabilistic Causality and the Question of
  Transitivity}.\BBCQ{}
\newblock{}\Bem{Philosophy of Science}, \Bem{50}, 35-57.

\bibitem[\protect\citeauthoryear{%
Einstein%
, Podolsky%
\BCBL{}\ \BBA{} Rosen%
}{%
Einstein%
\ \protect\BOthers{.}}{%
{\protect\APACyear{1935}}%
}]{%
epr}%
\APACinsertmetastar{%
epr}%
Einstein, A.%
, Podolsky, B.%
\BCBL{}\ \BBA{} Rosen, N.%
%
\newblock{}\BBOP{}1935\BBCP{}.
\newblock{}\BBOQ{}{Can quantum-mechanical description of physical reality be
  considered complete?}\BBCQ{}
\newblock{}\Bem{Physical Review}, \Bem{47}, 777--780.

\bibitem[\protect\citeauthoryear{%
Gisin%
}{%
Gisin%
}{%
{\protect\APACyear{1991}}%
}]{%
gisin91}%
\APACinsertmetastar{%
gisin91}%
Gisin, N.%
%
\newblock{}\BBOP{}1991\BBCP{}.
\newblock{}\BBOQ{}Bell's inequality holds for all non-product states.\BBCQ{}
\newblock{}\Bem{Physics Letters A}, \Bem{154}, 201-202.

\bibitem[\protect\citeauthoryear{%
Gisin%
}{%
Gisin%
}{%
{\protect\APACyear{2005}}%
}]{%
gisin05}%
\APACinsertmetastar{%
gisin05}%
Gisin, N.%
%
\newblock{}\BBOP{}2005\BBCP{}.
\newblock{}\Bem{{Can relativity be considered complete? From Newtonian
  nonlocality to quantum nonlocality and beyond}.}
\newblock{}(URL = http://arxiv.org/abs/quant-ph/0512168)

\bibitem[\protect\citeauthoryear{%
Gra{\ss}hoff%
, Portmann%
\BCBL{}\ \BBA{} W\"uthrich%
}{%
Gra{\ss}hoff%
\ \protect\BOthers{.}}{%
{\protect\APACyear{2005}}%
}]{%
grasshoff05}%
\APACinsertmetastar{%
grasshoff05}%
Gra{\ss}hoff, G.%
, Portmann, S.%
\BCBL{}\ \BBA{} W\"uthrich, A.%
%
\newblock{}\BBOP{}2005\BBCP{}.
\newblock{}\BBOQ{}{Minimal Assumption Derivation of a Bell-type
  inequality}.\BBCQ{}
\newblock{}\Bem{British Journal for the Philosophy of Science}, \Bem{56},
  663-680.
\newblock{}(URL = http://lanl.arxiv.org/abs/quant-ph/0312176)

\bibitem[\protect\citeauthoryear{%
Greenberger%
, Horne%
\BCBL{}\ \BBA{} Zeilinger%
}{%
Greenberger%
\ \protect\BOthers{.}}{%
{\protect\APACyear{1989}}%
}]{%
greenberger89}%
\APACinsertmetastar{%
greenberger89}%
Greenberger, D.%
, Horne, M.%
\BCBL{}\ \BBA{} Zeilinger, A.%
%
\newblock{}\BBOP{}1989\BBCP{}.
\newblock{}\BBOQ{}{Going beyond Bell's theorem}.\BBCQ{}
\newblock{}\BIn{} M.~Kafatos\ (\BED), \Bem{Bell's theorem, quantum theory, and
  conceptions of the universe}\ (\BPG\ 73-76).
\newblock{}Dordrecht: Kluwer.

\bibitem[\protect\citeauthoryear{%
Henson%
}{%
Henson%
}{%
{\protect\APACyear{2005}}%
}]{%
henson05}%
\APACinsertmetastar{%
henson05}%
Henson, J.%
%
\newblock{}\BBOP{}2005\BBCP{}.
\newblock{}\BBOQ{}Comparing causality principles.\BBCQ{}
\newblock{}\Bem{Studies In History and Philosophy of Modern Physics},
  \Bem{36}(3), 519-543.
\newblock{}(URL = http://xxx.lanl.gov/abs/quant-ph/0410051)

\bibitem[\protect\citeauthoryear{%
Hofer-Szab\'o%
}{%
Hofer-Szab\'o%
}{%
{\protect\APACyear{2006}}%
}]{%
hofer06}%
\APACinsertmetastar{%
hofer06}%
Hofer-Szab\'o, G.%
%
\newblock{}\BBOP{}2006\BBCP{}.
\newblock{}\Bem{{Separate- versus \emph{common}-common-cause-type derivations
  of the Bell inequalities}.}
\newblock{}(forthcoming)

\bibitem[\protect\citeauthoryear{%
Hofer-Szab\'o%
\ \BBA{} R\'edei%
}{%
Hofer-Szab\'o%
\ \BBA{} R\'edei%
}{%
{\protect\APACyear{2004}}%
}]{%
hofer04}%
\APACinsertmetastar{%
hofer04}%
Hofer-Szab\'o, G.%
\BCBT{}\ \BBA{} R\'edei, M.%
%
\newblock{}\BBOP{}2004\BBCP{}.
\newblock{}\BBOQ{}{Reichenbachian Common Cause Systems}.\BBCQ{}
\newblock{}\Bem{British Journal for the Philosophy of Science}, \Bem{43},
  1819-1826.
\newblock{}(URL = http://philsci-archive.pitt.edu/archive/00001246/)

\bibitem[\protect\citeauthoryear{%
Hofer-Szab\'o%
, R\'edei%
\BCBL{}\ \BBA{} Szab\'o%
}{%
Hofer-Szab\'o%
\ \protect\BOthers{.}}{%
{\protect\APACyear{1999}}%
}]{%
redei99}%
\APACinsertmetastar{%
redei99}%
Hofer-Szab\'o, G.%
, R\'edei, M.%
\BCBL{}\ \BBA{} Szab\'o, L.~E.%
%
\newblock{}\BBOP{}1999\BBCP{}.
\newblock{}\BBOQ{}{On Reichenbach's Common Cause Principle and Reichenbach's
  Notion of Common Cause}.\BBCQ{}
\newblock{}\Bem{British Journal for the Philosophy of Science}, \Bem{50}(3),
  377--399.
\newblock{}(URL = http://xxx.lanl.gov/abs/quant-ph/9805066)

\bibitem[\protect\citeauthoryear{%
Jarrett%
}{%
Jarrett%
}{%
{\protect\APACyear{1984}}%
}]{%
jarrett84}%
\APACinsertmetastar{%
jarrett84}%
Jarrett, J.~P.%
%
\newblock{}\BBOP{}1984\BBCP{}.
\newblock{}\BBOQ{}{On the Physical Significance of the Locality Conditions in
  the Bell Arguments}.\BBCQ{}
\newblock{}\Bem{No\^us}, \Bem{18}, 569--589.

\bibitem[\protect\citeauthoryear{%
Jones%
\ \BBA{} Clifton%
}{%
Jones%
\ \BBA{} Clifton%
}{%
{\protect\APACyear{1993}}%
}]{%
jones93}%
\APACinsertmetastar{%
jones93}%
Jones, M.%
\BCBT{}\ \BBA{} Clifton, R.%
%
\newblock{}\BBOP{}1993\BBCP{}.
\newblock{}\BBOQ{}{Against Experimental Metaphysics}.\BBCQ{}
\newblock{}\BIn{} P.~French, J.~T.E.~Euling\BCBL{}\ \BBA{} H.~Wettstein\
  (\BEDS), \Bem{Mid-west studies in philosophy}\ (\BVOL\ XVIII, \BPG\ 295-316).
\newblock{}Notre Dame: University of Notre Dame Press.

\bibitem[\protect\citeauthoryear{%
Mattingly%
}{%
Mattingly%
}{%
{\protect\APACyear{2005}}%
}]{%
mattingly05}%
\APACinsertmetastar{%
mattingly05}%
Mattingly, D.%
%
\newblock{}\BBOP{}2005\BBCP{}.
\newblock{}\BBOQ{}{Modern Tests of Lorentz Invariance}.\BBCQ{}
\newblock{}\Bem{Living Rev. Relativity}, \Bem{8}.
\newblock{}(URL (cited on 9 February 2006) =
  http://www.livingreviews.org/lrr-2005-5)

\bibitem[\protect\citeauthoryear{%
Maudlin%
}{%
Maudlin%
}{%
{\protect\APACyear{1994}}%
}]{%
maudlin94}%
\APACinsertmetastar{%
maudlin94}%
Maudlin, T.%
%
\newblock{}\BBOP{}1994\BBCP{}.
\newblock{}\Bem{Quantum non-locality and relativity}.
\newblock{}Cambridge: Blackwell.

\bibitem[\protect\citeauthoryear{%
Pearl%
}{%
Pearl%
}{%
{\protect\APACyear{2000}}%
}]{%
pearl00}%
\APACinsertmetastar{%
pearl00}%
Pearl, J.%
%
\newblock{}\BBOP{}2000\BBCP{}.
\newblock{}\Bem{Causality: Models, reasoning, and inference}.
\newblock{}New York: Cambridge University Press.

\bibitem[\protect\citeauthoryear{%
Pitowsky%
}{%
Pitowsky%
}{%
{\protect\APACyear{1989}}%
}]{%
pitowsky89}%
\APACinsertmetastar{%
pitowsky89}%
Pitowsky, I.%
%
\newblock{}\BBOP{}1989\BBCP{}.
\newblock{}\Bem{Quantum probability - quantum logic}\ (\BNUM\ 321).
\newblock{}Berlin: Springer-Verlag.

\bibitem[\protect\citeauthoryear{%
Redhead%
}{%
Redhead%
}{%
{\protect\APACyear{1987}}%
}]{%
redhead87}%
\APACinsertmetastar{%
redhead87}%
Redhead, M.%
%
\newblock{}\BBOP{}1987\BBCP{}.
\newblock{}\Bem{Incompleteness, nonlocality and realism}.
\newblock{}Clarendon Press.

\bibitem[\protect\citeauthoryear{%
Reichenbach%
}{%
Reichenbach%
}{%
{\protect\APACyear{1956}}%
}]{%
reichenbach56}%
\APACinsertmetastar{%
reichenbach56}%
Reichenbach, H.%
%
\newblock{}\BBOP{}1956\BBCP{}.
\newblock{}\Bem{The direction of time}.
\newblock{}Los Angeles: University of California Press.

\bibitem[\protect\citeauthoryear{%
Ryff%
}{%
Ryff%
}{%
{\protect\APACyear{1997}}%
}]{%
ryff97}%
\APACinsertmetastar{%
ryff97}%
Ryff, L.~C.%
%
\newblock{}\BBOP{}1997, December\BBCP{}.
\newblock{}\BBOQ{}{Bell and Greenberger, Horne, and Zeilinger theorems
  revisited}.\BBCQ{}
\newblock{}\Bem{American Journal of Physics}, \Bem{65}(12), 1197--1199.

\bibitem[\protect\citeauthoryear{%
Shimony%
}{%
Shimony%
}{%
{\protect\APACyear{2005}}%
}]{%
shimony05}%
\APACinsertmetastar{%
shimony05}%
Shimony, A.%
%
\newblock{}\BBOP{}2005\BBCP{}.
\newblock{}\BBOQ{}{Bell's Theorem}.\BBCQ{}
\newblock{}\BIn{} E.~N. Zalta\ (\BED), \Bem{The stanford encyclopedia of
  philosophy (summer 2005 edition).}
\newblock{}
\newblock{}(URL =
  http://plato.stanford.edu/archives/sum2005/entries/bell-theorem)

\bibitem[\protect\citeauthoryear{%
Skyrms%
}{%
Skyrms%
}{%
{\protect\APACyear{1980}}%
}]{%
skyrms80}%
\APACinsertmetastar{%
skyrms80}%
Skyrms, B.%
%
\newblock{}\BBOP{}1980\BBCP{}.
\newblock{}\Bem{Causal necessity}.
\newblock{}New Haven: Yale University Press.

\bibitem[\protect\citeauthoryear{%
Spirtes%
, Glymour%
\BCBL{}\ \BBA{} Scheines%
}{%
Spirtes%
\ \protect\BOthers{.}}{%
{\protect\APACyear{1993}}%
}]{%
spirtes93}%
\APACinsertmetastar{%
spirtes93}%
Spirtes, P.%
, Glymour, C.%
\BCBL{}\ \BBA{} Scheines, R.%
%
\newblock{}\BBOP{}1993\BBCP{}.
\newblock{}\Bem{Causation, prediction, and search}.
\newblock{}Berlin: Springer-Verlag.

\bibitem[\protect\citeauthoryear{%
Suppes%
\ \BBA{} Zanotti%
}{%
Suppes%
\ \BBA{} Zanotti%
}{%
{\protect\APACyear{1976}}%
}]{%
suppes76}%
\APACinsertmetastar{%
suppes76}%
Suppes, P.%
\BCBT{}\ \BBA{} Zanotti, M.%
%
\newblock{}\BBOP{}1976\BBCP{}.
\newblock{}\BBOQ{}{On the Determinism of Hidden Variable Theories with Strict
  Correlation and Conditional Statistical Independence of Observables}.\BBCQ{}
\newblock{}\BIn{} P.~Suppes\ (\BED), \Bem{Logic and probability in quantum
  mechanics}\ (\BPG\ 445-455).
\newblock{}Dordrecht: Reidel.

\bibitem[\protect\citeauthoryear{%
{Tsirelson (Cirel'son)}%
}{%
{Tsirelson (Cirel'son)}%
}{%
{\protect\APACyear{1980}}%
}]{%
tsirelson80}%
\APACinsertmetastar{%
tsirelson80}%
{Tsirelson (Cirel'son)}, B.%
%
\newblock{}\BBOP{}1980\BBCP{}.
\newblock{}\BBOQ{}{Quantum Generalizations of Bell's Inequality}.\BBCQ{}
\newblock{}\Bem{Letters in Mathematical Physics}, \Bem{4}, 93-100.

\bibitem[\protect\citeauthoryear{%
Uffink%
}{%
Uffink%
}{%
{\protect\APACyear{1999}}%
}]{%
uffink99}%
\APACinsertmetastar{%
uffink99}%
Uffink, J.%
%
\newblock{}\BBOP{}1999\BBCP{}.
\newblock{}\BBOQ{}{The Principle of the Common Cause Faces the Bernstein
  Paradox}.\BBCQ{}
\newblock{}\Bem{Philosophy of Science}(66), 512-525.

\bibitem[\protect\citeauthoryear{%
{van Fraassen}%
}{%
{van Fraassen}%
}{%
{\protect\APACyear{1982}}%
}]{%
vanfraassen82}%
\APACinsertmetastar{%
vanfraassen82}%
{van Fraassen}, B.~C.%
%
\newblock{}\BBOP{}1982\BBCP{}.
\newblock{}\BBOQ{}{The Charybdis of Realism: Epistemological Implications of
  Bell's Inequalities}.\BBCQ{}
\newblock{}\Bem{Synthese}, \Bem{52}, 25-38.

\bibitem[\protect\citeauthoryear{%
Weinstein%
}{%
Weinstein%
}{%
{\protect\APACyear{2006}}%
}]{%
weinstein06}%
\APACinsertmetastar{%
weinstein06}%
Weinstein, S.%
%
\newblock{}\BBOP{}2006\BBCP{}.
\newblock{}\BBOQ{}{Superluminal Signaling and Relativity}.\BBCQ{}
\newblock{}\Bem{Synthese}, \Bem{148}, 381-399.

\bibitem[\protect\citeauthoryear{%
W\"uthrich%
}{%
W\"uthrich%
}{%
{\protect\APACyear{2004}}%
}]{%
wuethrich04}%
\APACinsertmetastar{%
wuethrich04}%
W\"uthrich, A.%
%
\newblock{}\BBOP{}2004\BBCP{}.
\newblock{}\Bem{Quantum correlations and common causes}.
\newblock{}Bern: Bern Studies in the History and Philosophy of Science.

\end{thebibliography}

\end{document}